\documentclass[aps,prl,twocolumn,superscriptaddress]{revtex4-2}

\usepackage{amsmath, amssymb}
\usepackage{graphicx}
\usepackage[colorlinks = true, urlcolor = cyan, linkcolor=blue, citecolor=blue, filecolor=magenta]{hyperref}

\newcommand{\bra}[1]{\langle #1|}
\newcommand{\ket}[1]{|#1\rangle}

\begin{document}

\title{Exact Fock-State Preparation with $n^{1/4}$ Circuit Depth}

\author{Tanay Roy}
\email{roytanay@fnal.gov}
\affiliation{Superconducting Quantum Materials and Systems (SQMS) Center, Fermi National Accelerator Laboratory, Batavia, IL 60510, USA}

\date{\today}

\begin{abstract}

Efficient, deterministic, and high-fidelity preparation of large Fock states is essential for scaling bosonic quantum technologies and exploring quantum phenomena at large excitation energies. We introduce a deterministic one-parameter (D1p) protocol that maps Fock-state preparation in an infinite-dimensional Hilbert space onto two-dimensional amplitude amplification. Starting from a coherent state with $|\alpha|\simeq\sqrt{n}$, the initial target-state population scales as $n^{-1/2}$, yielding an iteration count and circuit depth of $\mathcal{O}(n^{1/4})$. Phase matching guarantees unit fidelity in the ideal model; remarkably, preparing $\ket{10^6}$ requires only 39 iterations. The protocol uses only displacements and number-selective phase operations, requires no numerical optimization, and further extends to state transfer, general superpositions, finite-dimensional systems, and multipartite entangled states. In the large-amplitude regime, its multi-target form prepares $L$-legged cat states with an iteration count determined only by $L$; cats with up to ten legs require only two iterations, independent of the coherent-state amplitude. This framework provides a broadly applicable route to highly excited bosonic states on platforms supporting these elementary controls.

\end{abstract}

\maketitle


\textit{Introduction}---Bosonic modes underpin quantum-information processing across cavity and circuit quantum electrodynamics (QED)~\cite{Blais2021cqed}, trapped ions~\cite{Katz2023trapped_ion}, optomechanics~\cite{Aspelmeyer2014optomech}, and magnonics~\cite{Zare2022magnon}. Their energy eigenstates---Fock states---and arbitrary superpositions are essential non-Gaussian resources for quantum computation~\cite{Michael2016binomial}, communication~\cite{Hilmann2022teleport}, metrology~\cite{Deng2024metrology}, and sensing~\cite{Chen2025sensing}. Extending Fock-state preparation to large photon numbers is therefore of profound importance, both for exploring novel quantum phenomena at macroscopic scales~\cite{Frowis2018macro} and for enabling advances in quantum simulation~\cite{Saugmann2023lattice, Deng2022topology, Sturges2021sim}, quantum algorithms~\cite{Gan2022ML, Wang2017boson, Zhong2020advantage}, and quantum-enhanced sensing~\cite{Deng2024metrology, Agrawal2024DM}.

Despite substantial progress in cavity and circuit QED~\cite{laweberly1996universal, Varcoe2000fock, Brattke2001fock, Uria2020fock, Meher2026fock, Hofheinz2008fock, Gevorgyan2012fock_kerr, Premaratne2017fock, Gu2017fock}, trapped-ion~\cite{Cirac1993fock_ion, Meekhof1996fock_ion, Wolf2019fock_ion}, phononic~\cite{Chu2018fock_phonon}, and optical platforms~\cite{Gonzalez2015fock_optical, Tiedau2019fock_optical, Cooper2013fock_optical}, preparation at macroscopic occupation remains challenging. Postselected approaches suffer rapidly decreasing success probabilities~\cite{Tiedau2019fock_optical, Sayrin2011fock_meas, Snaka2005fock_meas, Thekkadath2020fock_meas, Deng2024metrology}, whereas deterministic schemes may sacrifice fidelity or require control complexity that grows with Hilbert-space dimension~\cite{Brattke2001fock, engelkemeier2021fock_optical}. Universal and optimal-control based methods often demand long pulse sequences and numerical optimization with control complexity increasing rapidly with photon number~\cite{SNAP2015PRL, SNAP2015PRA, Sivak2022reinforcement, huang2026universal_qudit}. Recent Hamiltonian-level proposals reduce some control complexity but still rely on numerical optimization and resonator nonlinearity~\cite{li2026fock_10k, Xu2026fock}. Meanwhile, environmental coupling produces an effective decay rate that grows linearly with occupation, so preparation must become faster rather than merely more accurate as $n$ increases. Experiments therefore remain limited to $\lesssim100$ photons~\cite{kim2025ultracoherent, Deng2024metrology}, motivating analytical protocols with sublinear runtime, modest Hilbert-space control, and experimentally simple operations.


We address these challenges by introducing the deterministic one-parameter (D1p) protocol, which maps the dynamics in the infinite-dimensional Hilbert space onto the two-dimensional subspace defined by the target and an initial state with overlap $\lambda$. Within this invariant subspace, Fock-state preparation becomes a Grover-search problem rather than a task requiring independent control of every populated Fock level. The resulting amplitude-amplification circuit has depth $\mathcal{O}(1/\sqrt{\lambda})$, the optimal Grover scaling under the available oracle model. Preparing $\ket n$ from $\ket{\alpha=\sqrt n}$ therefore requires gate count and runtime $\mathcal{O}(n^{1/4})$. A recent proposal~\cite{jin2026fock_RSE} similarly confines the infinite-dimensional dynamics to a two-dimensional subspace and derives analytical conditions for continuous-time evolution. Its actual discrete implementation, however, requires numerical optimization of two phase angles per iteration. D1p instead applies one matched phase to both the oracle and diffusion operators---hence ``one parameter"---and determines this phase and the optimal iteration count in closed form for arbitrary $n$. It therefore requires no numerical optimization and achieves unit fidelity in the ideal model. We further extend the construction to state transfer, selected superpositions, finite-dimensional systems, and multipartite entangled states.

\begin{figure*}[t]
    \centering
    \includegraphics[width=\textwidth]{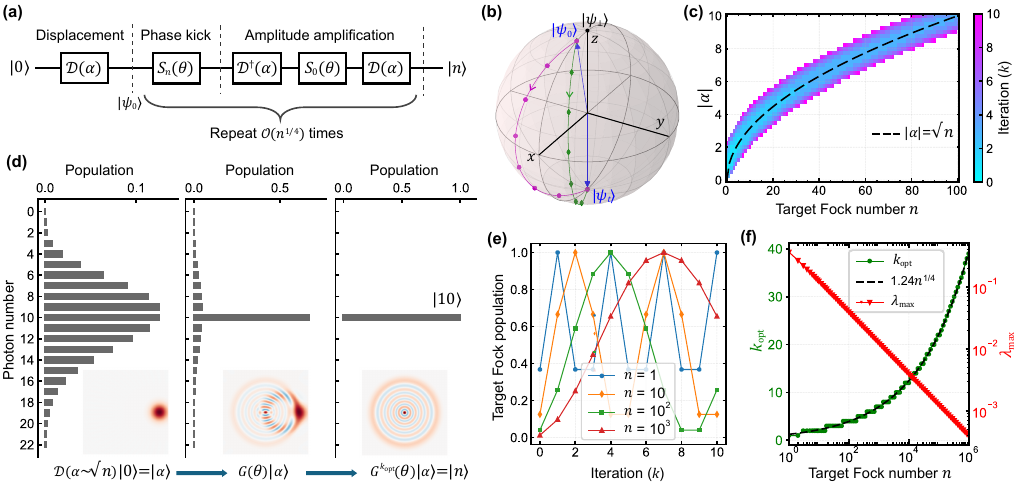}
    \caption{\textbf{Deterministic Fock-state preparation.} (a) Circuit based on phase kicks and amplitude amplification; the bracketed gates form one iterate $G(\theta)=S_r(\theta)S_o(\theta)$. (b) Trajectories from $\ket{\alpha=\sqrt{1000}}$ to $\ket{1000}$ in seven iterations for positive (green) and negative (purple) phase kicks, with $|\theta|=2.393$. (c) Iteration count versus target $n$ and initial coherent amplitude $\alpha$; the dashed line marks the optimum. (d) Fock populations (Wigner tomographs in inset) during preparation of $\ket{10}$. (e) Target population versus iteration count showing $k_{\rm opt} = \{1,2,4,7 \}$ for $n = \{1,10,100,1000 \}$, respectively. (f) Maximum overlap $\lambda_{\rm max}$ and optimal iteration count $k_{\rm opt}$ versus $n$, with $k_{\rm opt}\approx 1.24n^{1/4}$.}
    \label{fig:fig1}
\end{figure*}

\textit{Amplitude amplification}---We first summarize the amplitude-amplification framework that underlies the preparation protocol. It combines two operations: an oracle $S_o$ that changes the phase of the target state $\ket{\psi_t}$ and a diffusion operator $S_r$ that rotates the state about the axis defined by the initial state $\ket{\psi_0}$. Generalized phase rotations, rather than only reflections, are described by
\begin{equation}
\begin{split}
    S_o(\phi) = \mathbb{I} - \left(1-e^{i\phi}\right) \ket{\psi_t} \bra{\psi_t}, \\
    S_r(\phi) = \mathbb{I} - \left(1-e^{i\phi}\right) \ket{\psi_0} \bra{\psi_0},
\end{split}
\end{equation}
and one Grover iterate is $G(\phi)=S_r(\phi)S_o(\phi)$. Repeated application changes only the relative amplitudes in the subspace spanned by the target and its orthogonal complement $\ket{\psi_t}$. The minimum number of iterations $k_{\rm opt}$ to reach $\ket{\psi_t}$ exactly depends on the nonzero initial overlap $\lambda=|\langle\psi_t|\psi_0\rangle|^2$ and is given by
\begin{equation}
\label{eq:kopt}
k_{\rm opt} = \left \lceil \dfrac{\pi}{4\sin^{-1}\sqrt{\lambda}} - \dfrac{1}{2} \right \rceil,
\end{equation}
showing a larger overlap reduces $k_{\rm opt}$. Here $\lceil\cdot\rceil$ denotes the ceiling function. The conventional choice $\phi=\pi$ approaches, but does not generally reach, unit target probability~\cite{Grover_algo}, because the fixed-angle rotation can undershoot or overshoot the target~\cite{Roy2022D2p}. We instead use the deterministic phase-matching prescription of Long \textit{et al.}~\cite{Long2001certain_Grover, Roy2026qudit_Grover}. Because the same matched phase $\phi=\theta$ is used for both rotations in every iterate $G(\theta)$, we call this construction the deterministic one-parameter (D1p) protocol with
\begin{equation}
\label{eq:d1p_theta}
    \theta = \pm 2 \sin^{-1} \left( \frac{1}{\sqrt{\lambda}} \sin{\left(\frac{\pi}{4k_{\rm opt}+2} \right)} \right),
\end{equation}
The two signs generate trajectories on opposite sides of the effective Bloch sphere spanned by $\ket{\psi_t}$ and $\ket{\psi_\perp}$ but give the same population dynamics (see Fig.~\ref{fig:fig1}(b)). Although Grover search is usually formulated in finite dimension, its reduction to a two-dimensional invariant subspace does not depend on the dimension of the ambient Hilbert space.

\textit{Deterministic Fock-state preparation}---We now specialize this framework to a harmonic oscillator and show that both required rotations can be synthesized from native bosonic controls. Consider a linear bosonic system with the Hamiltonian
\begin{equation}
    H_0/\hbar = \omega_r \hat{a}^\dagger \hat{a} + \Omega (\hat{a} + \hat{a}^\dagger) \sin{(\omega_d t)},
\end{equation}
where $\omega_r$ and $\omega_d$ are the resonator and drive frequencies, $\Omega$ is the drive amplitude, and $\hat a$ is the annihilation operator. On resonance, the drive realizes a displacement whose magnitude is proportional to the pulse area and whose phase is set by the drive phase.

To prepare $\ket n$ from $\ket0$, we first create a coherent state $\ket{\psi_0}=\ket\alpha$ with the displacement $\mathcal D(\alpha)=e^{\alpha\hat a^\dagger-\alpha^*\hat a}$. The Poisson weight of the target is
\begin{equation}
\label{eq:lambda_n}
    \lambda = |\bra{\psi_0}n\rangle|^2 = \frac{e^{-|\alpha|^2} |\alpha|^{2n}}{{n!}}.
\end{equation}
The coherent state has nonzero support on every Fock level, so any finite target $\ket n$ is accessible. Although the drive initially populates many levels, the oracle and diffusion operators preserve the two-dimensional subspace spanned by $\ket{\psi_t}=\ket n$ and
\begin{equation}
    \ket{\psi_\perp} = \frac{\ket{\psi_0}-\eta \ket{\psi_t}}{\sqrt{1-|\eta|^2}},
\end{equation}
where $\eta=\langle\psi_t|\psi_0\rangle$; see Fig.~\ref{fig:fig1}(b). Thus the infinite set of nontarget Fock components enters only through the single normalized vector $\ket{\psi_\perp}$. This reduction is the central reason that the protocol does not require level-by-level pulse optimization.

Both operators can be constructed from native controls in circuit QED. The oracle is the number-selective phase kick $\ket n\mapsto e^{i\phi}\ket n$, a single-component selective number-dependent arbitrary phase (SNAP) gate $S_n(\phi)$~\cite{SNAP2015PRL}; analogous number-selective phases are available in trapped ions~\cite{Park2024qsim}. Importantly, this oracle addresses only the target level and does not require a general SNAP gate with independently optimized phases on the entire occupied manifold. For $\ket{\psi_0}=\ket\alpha = \mathcal D(\alpha)\ket0$, the diffusion operator is obtained by mapping the coherent state back to vacuum, applying $S_0(\phi)$, and undoing that mapping:
\begin{equation}
    S_r(\phi)=\mathcal D(\alpha)S_0(\phi)\mathcal D^\dagger(\alpha).
\end{equation}
Figure~\ref{fig:fig1}(a) shows the full sequence: initial displacement, oracle phase kick, and repeated diffusion--oracle pairs. Since the same displacement amplitude and phase kick are reused, the control description remains simple as $n$ grows. The protocol is deterministic for any $\lambda>0$ corresponding to different $|\alpha|$ as shown in Fig.~\ref{fig:fig1}(c), but maximizing $\lambda$ minimizes circuit depth. For fixed $n$, Eq.~\eqref{eq:lambda_n} is maximized at $|\alpha|=\sqrt n$, giving
\begin{equation}
    \lambda_{\rm max}= \frac{e^{-n} n^n}{n!} \approx \frac{1}{\sqrt{2\pi n}},
\end{equation}
where we used Stirling's approximation $n!\simeq\sqrt{2\pi n}(n/e)^n$. The best coherent seed therefore has a target probability that decreases only as $n^{-1/2}$ rather than exponentially. For small $\lambda$, Eq.~\eqref{eq:kopt} gives
\begin{equation}
\label{eq:n_scaling}
k_{\rm opt} \approx \left \lceil \dfrac{\pi}{4\sqrt{\lambda}} - \dfrac{1}{2} \right \rceil = \left \lceil \dfrac{\pi}{4}(2\pi n)^{1/4} - \dfrac{1}{2} \right \rceil,
\end{equation}
resulting in $k_{\rm opt}$ scaling as $\mathcal O(n^{1/4})$, as shown in Fig.~\ref{fig:fig1}(f). The leading coefficient is approximately $1.24$, so even preparing $\ket{10^6}$ requires only 39 iterations. If individual gate durations remain approximately independent of $n$, the preparation time inherits the same sublinear scaling as the circuit depth. Other deterministic amplitude-amplification protocols may also be used without changing this asymptotic scaling~\cite{Roy2022D2p, Roy2026qudit_Grover}.

Figure~\ref{fig:fig1}(d) illustrates the state evolution when preparing $\ket{10}$ from $\ket{\alpha=\sqrt{10}}$ and \ref{fig:fig1}(e) shows population dynamics of the target $\ket{n}$ for a few representative values. Each iterate coherently removes amplitude from every nontarget component and transfers it to the target. The target reaches unit population at the level-dependent iteration count $k_{\rm opt}$; further iterations reverse the transfer and produce the familiar Grover oscillation.

The protocol remains efficient over a broad range of $\alpha$ [Fig.~\ref{fig:fig1}(c)], so experiments need not set $\alpha=\sqrt n$ precisely. Values near this optimum often yield the same integer $k_{\rm opt}$ even though the matched phase $\theta$ changes slightly. At the population maximum, the success probability is also first-order insensitive to small errors in both $\alpha$ and $\theta$, providing robustness against slow drift and calibration fluctuations. 

Modern control schemes realize the SNAP phase virtually~\cite{SNAP2015PRL} or through pulse engineering~\cite{You2025floquet_snap} for faster operation where the phase resolution should not be a limiting factor. If phase matching is not used, the conventional choice $\phi=\pi$ gives the nondeterministic fidelity $\mathcal F(k)=\sin^2[(2k+1)\sin^{-1}(\sqrt\lambda)]$, which approaches unity near $k_{\rm opt}$ but generally does not equal it. This fixed-phase variant trades exactness for the simplest possible phase calibration.

Further, the displacement phase is irrelevant for single-Fock-state preparation: $\alpha e^{i\varphi}$ may point in any phase-space direction because the overlap in Eq.~\eqref{eq:lambda_n} depends only on $|\alpha|$. This freedom can simplify pulse scheduling and avoid phase constraints imposed by other control tones.


\begin{figure*}[t]
    \centering
    \includegraphics[width=\textwidth]{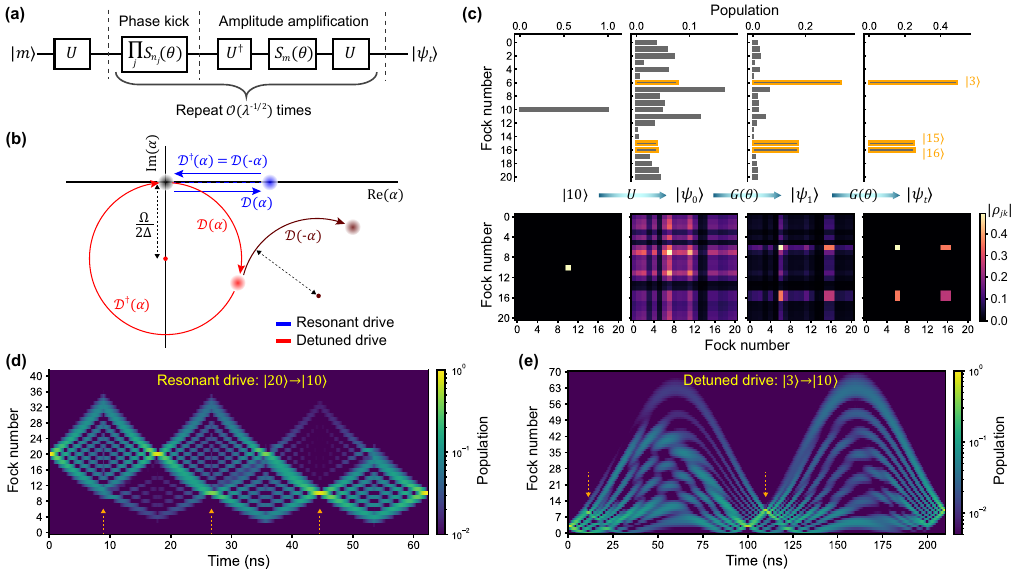}
    \caption{\textbf{Generalized protocol.} (a) Circuit for a finite- or infinite-dimensional system, with $\ket{\psi_0}=U\ket m$ and $\lambda=|\langle\psi_t|\psi_0\rangle|^2>0$. The oracle addresses multiple basis states. (b) Coherent-state trajectories under resonant (blue) and detuned (red) drives. (c) Fock populations (top panel) and density-matrix elements (bottom panel) for a 21-dimensional system starting from $\ket{10}$, with targets $\{\ket3,\ket{15},\ket{16}\}$, $\lambda=0.156$, and $k_{\rm opt}=2$. (d) Resonant-drive transfer with $\lambda=0.075$ and $\theta=1.900$; orange dashed lines mark instantaneous phase kicks. (e) Detuned-drive transfer with $\lambda=0.101$, $\Delta/2\pi=10~\mathrm{MHz}$, and $\theta=2.679$.}
    \label{fig:fig2}
\end{figure*}

\textit{Extensions and applications}---Having established the single-Fock-state protocol and its implementation, we next develop increasingly general applications: transfer between Fock states, preparation of selected and arbitrary superpositions, extension to finite-dimensional systems, and finally multipartite entanglement. Figure~\ref{fig:fig2}(a) summarizes the generalized circuit for an initial basis state $\ket m$, an arbitrary state-preparation unitary $U$, and an oracle acting on multiple basis states. In this setting $\ket{\psi_0}=U\ket m$, and the only mathematical requirement is a nonzero overlap with the target subspace. The diffusion operator is implemented as $US_m(\theta)U^\dagger$, so the selective phase acts on $\ket m$ rather than the usual vacuum state. This observation separates amplitude amplification from the specific displacement controls used for a harmonic oscillator.

\textit{(i) State transfer}---The most direct extension is transfer between two Fock states. One could concatenate inverse preparation of $\ket m$ with preparation of $\ket n$ through the path $\ket m\to\ket0\to\ket n$, but the two iteration counts would add. A shorter route uses the displaced initial state $\mathcal D(\alpha')\ket m$ and chooses $\alpha'$ to maximize $|\langle n|\mathcal D(\alpha')|m\rangle|$. This overlap is determined by a displaced-number-state matrix element and can be optimized as a one-dimensional function. Although the exact optimum depends on both indices, $|\sqrt m-\sqrt n|$ provides a useful scale when $m$ and $n$ are large and nearby. The displacement phase again does not affect the overlap magnitude. Consequently, transfers between nearby highly excited states can require substantially fewer iterations than preparing each state from vacuum. Figure~\ref{fig:fig2}(d) simulates $\ket{20}\to\ket{10}$ for $\omega_d=\omega_r=2\pi\times5.0~\mathrm{GHz}$ and $\Omega=2\pi\times50~\mathrm{MHz}$. The transfer requires three iterations at $\alpha'=1.4$, corresponding to an $8.91$-ns displacement.

\textit{(ii) Detuned-drive implementation}---The state-transfer protocol does not require resonant displacements. With a detuned drive, however, the inverse operation must account for the curved phase-space trajectory. For a resonant rectangular pulse of duration $t_{\rm pulse}$, $\alpha=\Omega t_{\rm pulse}/2$ and the trajectory is straight (in a frame rotating at $\omega_r$). Its direction is set by the microwave phase, so advancing that phase by $\pi$ realizes $\mathcal D^\dagger(\alpha)=\mathcal D(-\alpha)$. With detuning $\Delta=\omega_d-\omega_r$, an initial coherent state instead follows a circle of radius $|\Omega/(2\Delta)|$ centered away from its initial position. Its evolution from time $t_0$ to $t$ is
\begin{equation}
    \ket{\alpha(t,t_0)} = \ket{\alpha(t_0) + \frac{i\Omega}{2\Delta}\left( e^{-i\Delta t} - e^{-i\Delta t_0} \right)}.
\end{equation}
For $\Delta>0$ the trajectory rotates clockwise [Fig.~\ref{fig:fig2}(b)]. Unlike the resonant case, reversing the drive phase does not retrace the same arc because it changes the center of rotation. The inverse instead completes the original circle. Thus a pulse of duration $t_{\rm period}-t_{\rm pulse}$, where $t_{\rm period}=2\pi/|\Delta|$, realizes the inverse evolution up to the same rotating-frame convention. This longer inverse pulse increases the runtime but leaves the amplitude-amplification construction unchanged. Figure~\ref{fig:fig2}(e) shows transfer from $\ket3$ to $\ket{10}$ with $\Delta/2\pi=10~\mathrm{MHz}$ and $t_{\rm pulse}=10~\mathrm{ns}$.

\textit{(iii) Selected Fock superpositions}---Beyond transferring population to a single level, a multi-target oracle can amplify an entire selected subspace. Multi-target amplitude amplification prepares the normalized projection of the initial state onto chosen Fock levels. Let $\ket{\psi_0}=\sum_jc_j\ket j$. For selected levels $n_1,\ldots,n_M$, the oracle applies the same phase to every member of the target subspace:
\begin{equation}
    S_{o, \rm{multi}}(\theta) = \prod_{j=1}^{M} S_{n_j}(\theta) = \sum_{j=1}^{M} e^{i\theta} \ket{n_j}\bra{n_j}
\end{equation}
and $\lambda_M=\sum_{j=1}^M|c_{n_j}|^2$. The final state is
\begin{equation}
    \ket{\psi_t}=\frac{1}{\sqrt{\lambda_M}}\sum_{j=1}^M c_{n_j}\ket{n_j}
    =\frac{\hat P\ket{\psi_0}}{\sqrt{\lambda_M}},
\end{equation}
where $\hat P=\sum_{j=1}^M\ket{n_j}\bra{n_j}$ is the projection operator. Amplitude amplification changes the total weight in the target subspace but does not rotate states within it. Because SNAP gates are diagonal in the Fock basis, their order is immaterial, and the selected coefficients retain the relative amplitudes and phases inherited from the coherent state. The accessible superpositions are therefore restricted by the initial coefficient ratios, but their preparation still requires only a multi-component SNAP gate and the same diffusion operation.

For example, $|\alpha|=\sqrt n$ gives $|c_{n-1}|=|c_n|$, so addressing both levels prepares $(\ket{n-1}+e^{i\varphi}\ket n)/\sqrt2$. The displacement phase sets $\varphi$ because adjacent coherent-state coefficients differ by the phase of $\alpha$. Such states form a number-state interferometer with a one-quantum separation and are useful for dephasing measurements and quantum sensing.

\textit{(iv) Multi-legged cat states}---The multi-target construction above also yields a particularly efficient route to bosonic cat codes~\cite{Cochrane1999cat2}. An $L$-legged cat state with logical index $m$ is
\begin{equation}
|C_L^{(m)}\rangle = \mathcal{N}_L \sum_{j=0}^{L-1}
e^{-i\frac{2\pi j}{L}m}\,\big|\alpha e^{i \frac{2\pi j}{L}}\big\rangle,
\end{equation}
a superposition of $L$ coherent states equally spaced on a circle of radius $\alpha$. Seeding the protocol with one constituent coherent state,
$|\psi_0\rangle=|\alpha\rangle$, and expanding both
states in the Fock basis, the $L$-fold phase sum in
$|C_L^{(m)}\rangle$ projects onto Fock levels
$n=m+jL$ ($j=0,1,2,\ldots$), so that
\begin{equation}
\lambda = |\langle\alpha|C_L^{(m)}\rangle|^2 = P_m,
\,
P_m \equiv e^{-\alpha^2}\sum_{j=0}^\infty
\frac{\alpha^{2(m+jL)}}{(m+jL)!},
\end{equation}
exactly, where $P_m$ is the total Poissonian population of
$|\alpha\rangle$ on the residue class $n\equiv m
\pmod L$. For $\alpha^2\gg 1$ the Poisson distribution is
approximately equidistributed across the $L$ residue
classes, giving $\lambda\approx 1/L$ independent of
$\alpha$, so that distinct legs are approximately orthogonal and normalization constant $\mathcal{N}_L \approx 1/\sqrt{L}$. The optimal
iteration count $k_{\rm opt} \approx \lceil \pi/[4\sin^{-1}(1/\sqrt{L})] -1/2\rceil$ then depends only on the number of legs $L$: for example, cat states with up to $L=10$ legs can be prepared in only two iterations, independent of $\alpha$.

In the Fock basis, $|C_L^{(m)}\rangle$ is a superposition of levels $n = m + jL$ for all $j=0,1,2,\ldots$, so a naive multi-target oracle would need to address an infinite set of Fock components. In practice, however, the Fock-state weights inherited from the underlying coherent state follow a Poisson envelope peaked at $n\approx \alpha^2$ with width $\sim\alpha$, so only the finite set of levels $n=m+jL$ falling within this envelope carry non-negligible population. The multi-target oracle can therefore be truncated to this finite window without loss of fidelity, keeping the construction experimentally finite-depth even though the logical cat state formally spans infinitely many Fock levels.

\textit{(v) Arbitrary superposition states}---The preceding construction preserves the coefficient ratios inherited from the initial state. Removing this restriction requires a state-selective oracle. An arbitrary superposition is possible in principle for $\lambda>0$, but requires $S_o(\theta)=\mathbb I-(1-e^{i\theta})\ket{\psi_t}\bra{\psi_t}$ rather than a common phase on several basis states. The distinction is important: a diagonal multi-target oracle amplifies a subspace, whereas the state-selective oracle identifies one direction within that subspace. SNAP and displacement gates form a universal control set~\cite{SNAP2015PRA, SNAP2015PRL}, so the latter can be compiled, but generally not as a single native operation and possibly with a depth that grows with the number of components. Likewise, an arbitrary initial state is formally allowed, but its associated diffusion operator may be costly to realize. The efficiency advantage is therefore greatest when both operators have compact native decompositions.

\textit{(vi) Finite-dimensional systems}---The amplitude-amplification principle is not restricted to bosonic modes. It applies to any finite-dimensional system whenever a unitary $U$ prepares an initial state with nonzero target overlap. Displacements are then replaced by $U$ and $U^\dagger$, while basis-selective phases implement the oracle and diffusion operations. In Fig.~\ref{fig:fig2}(c), a random $U$ (with seed = 2) acts in 21 dimensions and the oracle selects $\ket3$, $\ket{15}$, and $\ket{16}$. Their combined initial weight is $\lambda=0.156$, giving $k_{\rm opt}=2$. After two iterations, amplitude amplification eliminates the nontarget components while preserving the relative target amplitudes.

\begin{figure}[t]
    \centering
    \includegraphics[width=\columnwidth]{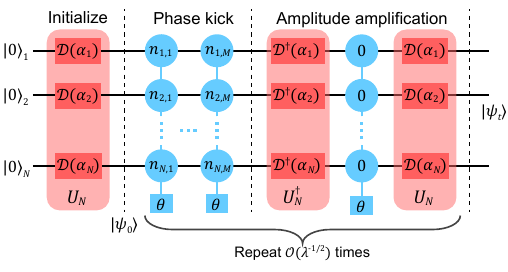}
    \caption{\textbf{Multipartite-state preparation.} Local displacements or a general unitary prepare $\ket{\psi_0}$ in $N$ bosonic modes or qudits. Generalized controlled-phase gates address the target components, while the diffusion step applies a controlled phase to $\ket0^{\otimes N}$.}
    \label{fig:fig3}
\end{figure}

\textit{(vii) Multipartite entangled states}---Finally, the construction extends from a single system to a composite Hilbert space. For $N$ modes initialized in $\ket0^{\otimes N}$, local displacements or a general unitary $U_N$ prepare a product or entangled state with overlap $\lambda$ with the desired target. Joint number-selective phase gates then implement the oracle on the target basis components, while a controlled phase on $\ket0^{\otimes N}$ generates the diffusion step after conjugation by the preparation unitary (Fig.~\ref{fig:fig3}). The same two-dimensional reduction applies because the oracle distinguishes only the target direction from its complement. The construction is therefore independent of whether the constituent systems are finite-dimensional qudits or bosonic modes, although the cost of joint controlled phases will depend strongly on the hardware connectivity.

\textit{Conclusion}---In summary, we have recast Fock-state preparation as exact amplitude amplification in a two-dimensional invariant subspace. For the optimal coherent seed $\ket{\alpha \simeq \sqrt n}$, the iteration count scales as $n^{1/4}$, while phase matching guarantees unit fidelity in the ideal model. The multi-target construction also provides an efficient route to multi-legged cat states: in the large-amplitude regime, the iteration count depends only on the number of legs, and cats with up to ten legs require only two iterations regardless of coherent-state amplitude. The explicit circuit uses only displacements and number-selective phase gates, requires no numerical search over the full Hilbert space, and further extends to state transfer, general superpositions, finite-dimensional systems, and multipartite targets. Its practical reach is set chiefly by number-selective control, critical-photon constraints, and occupation-enhanced loss. By separating the analytical design of the state-preparation protocol from hardware-specific gate optimization, D1p provides a scalable and platform-independent route toward highly excited nonclassical states.

\begin{acknowledgments}
\textit{Acknowledgments}---This work was supported by the U.S. Department of Energy, Office of Science, National Quantum Information Science Research Centers, Superconducting Quantum Materials and Systems Center (SQMS), under Contract No. 89243024CSC000002. Fermilab is operated by Fermi Forward Discovery Group, LLC under Contract No. 89243024CSC000002 with the U.S. Department of Energy, Office of Science, Office of High Energy Physics.
\end{acknowledgments}

\textit{Data availability}---No experimental datasets were generated or analyzed during this study. The results are based on analytical derivations and numerical calculations and are reproducible using the equations and methods described in the manuscript.

\bibliography{fock}

\clearpage
\onecolumngrid
\begin{center}
\section*{End Matter}
\end{center}
\twocolumngrid


\textit{Experimental considerations}---The ideal D1p construction is independent of a particular hardware platform. Its experimental performance is instead determined by the duration and fidelity of the displacement and number-selective phase operations. We discuss the principal considerations below, using circuit QED as a representative implementation.


\textit{(i) Scaling, runtime, and control precision}---The favorable $n^{1/4}$ scaling keeps the required number of logical operations small even at high occupation. The protocol requires only four iterations for $\ket{100}$ and 39 iterations for $\ket{10^6}$ [Fig.~\ref{fig:fig1}(f)]. Moreover, every iteration reuses the same displacement amplitudes and number-selective phase angles rather than independently controlling all populated Fock levels.

The physical runtime inherits this sublinear scaling. In circuit QED, displacements typically take tens of nanoseconds, whereas SNAP gates can take approximately $1~\mu\mathrm{s}$, depending on the resonator--qubit dispersive shift $\chi$. SNAP gates therefore dominate the sequence duration, but even the idealized $n=10^6$ sequence can remain below $100~\mu\mathrm{s}$. Larger dispersive shifts and error-detectable SNAP implementations can further improve performance without changing the logical protocol.

Phase precision should not impose a comparable limitation. The phase angle is programmed electronically, and modern control hardware resolves increments substantially smaller than those required by D1p. In addition, the target population is first-order insensitive to small errors in both $\alpha$ and $\theta$ at its maximum. These properties provide robustness against finite calibration precision and slow experimental drift.


\textit{(ii) Dispersive-control limitations}---In circuit QED, a SNAP gate is commonly generated by dispersively coupling the resonator to a nonlinear ancilla. This description assumes that the coupling is much smaller than the resonator--ancilla detuning and that the resonator occupation remains below the critical photon number. At higher occupation, corrections to the dispersive Hamiltonian shift the number-dependent transition frequencies, reduce selectivity, and can hybridize the resonator and ancilla. The accessible Fock number is therefore bounded by the hardware rather than by the amplitude-amplification protocol.

This limit can be modified through device design, for example by changing the coupling or detuning. Such changes generally introduce a tradeoff: increasing the critical photon number can reduce the dispersive shift and hence lengthen the SNAP gate. An implementation must therefore balance photon-number range, spectral selectivity, and gate duration.


\textit{(iii) Decoherence and coherent errors}---Relaxation and dephasing ultimately determine the preparation fidelity. Although high-quality cavities can have lifetimes of tens of milliseconds, the decay rate of a Fock state scales approximately as $n\kappa$, making highly excited states substantially shorter lived than the single-photon state. Ancilla relaxation or dephasing during a SNAP pulse can also imprint number-dependent errors on the resonator. A quantitative fidelity estimate must consequently integrate loss over the complete population trajectory and include ancilla errors, finite selectivity, and calibration errors.

Self-Kerr evolution provides an additional coherent error. For weakly nonlinear resonators satisfying $|\chi_s|\ll\Omega\ll\omega_r$, it is perturbative during an individual displacement, although the accumulated phase grows with occupation and sequence duration. For larger $\chi_s$, the Kerr evolution must be incorporated into the pulse design, compensated by calibrated phases, or explicitly refocused.

Finally, the simulations presented assume instantaneous, perfectly selective phase kicks. Finite-duration SNAP gates permit loss and unwanted Hamiltonian evolution during the oracle and may require shaped pulses or optimal control at the native-gate level. These nonidealities do not alter the ideal $n^{1/4}$ iteration scaling, but they set the achievable experimental fidelity. Thus D1p reduces the logical depth of Fock-state preparation without removing the requirement for high-fidelity native operations.





\end{document}